\documentclass[twocolumn,aps,pre,preprintnumbers,amsmath,amssymb,nofootinbib]{revtex4-1}

\usepackage[dvipdfmx]{graphicx}
\usepackage{dcolumn}
\usepackage{bm}
\usepackage{textcomp}
\usepackage{booktabs}
\usepackage{color}
\usepackage[T1]{fontenc}

\begin{document}

\title{Three-phase coexistence in binary charged lipid membranes in hypotonic solution} 

\author{Jingyu Guo$^1$, Hiroaki Ito$^2$, Yuji Higuchi$^3$, Klemen Bohinc$^4$, Naofumi Shimokawa$^{1*}$, and Masahiro Takagi$^1$} 
\affiliation
{$^1$School of Materials Science, Japan Advanced Institute of Science and Technology,
Ishikawa 923-1292, Japan \\
$^2$Department of Physics, Chiba University, Chiba 263-8522, Japan \\
$^3$Institute for Solid State Physics, University of Tokyo, Chiba, 227-8581, Japan \\
$^4$Faculty of Health Sciences, University of Ljubljana, SI-1000 Ljubljana, Slovenia}

\footnote[0]{$^*$nshimo@jaist.ac.jp}

\date{\today}

\begin{abstract}

We investigated the phase separation of dioleoylphosphatidylserine (DOPS) and dipalmitoylphosphatidylcholine (DPPC) in giant unilamellar vesicles in hypotonic solution using fluorescence and confocal laser scanning microscopy.
Although phase separation in charged lipid membranes is generally suppressed by the electrostatic repulsion between the charged headgroups, osmotic stress can promote the formation of charged lipid domains.
Interestingly, we observed three-phase coexistence even in DOPS/DPPC binary lipid mixtures.
The three phases were DPPC-rich, dissociated DOPS-rich, and nondissociated DOPS-rich phases.
The two forms of DOPS were found to coexist owing to the ionization of the DOPS headgroup, such that the system could be regarded as quasi-ternary.
The three formed phases with differently ionized DOPS domains were successfully identified experimentally by monitoring the adsorption of positively charged particles.
In addition, coarse-grained molecular dynamics simulations confirmed the stability of the three-phase coexistence.
Attraction mediated by hydrogen bonding between protonated DOPS molecules and reduction of the electrostatic interactions at the domain boundaries stabilized the three-phase coexistence.

\end{abstract}

\maketitle

\section{Introduction}
\label{intro}

Phospholipids, the main component of biomembranes, are amphiphilic molecules that self-assemble in aqueous solutions to form bilayer structures such as vesicles.
The physical properties of phospholipid bilayers have attracted considerable attention for understanding the physical phenomena and material properties of biomembranes.
In multicomponent lipid vesicles, the lipid species are distributed heterogeneously depending on the conditions, such as temperature and lipid composition, and phase-separated structures may also emerge~\cite{veatch02, veatch03, baumgart03, feigenson09, heberle11}.
It is believed that these phase-separated domains could serve as a model for the raft regions that may exist as compositionally heterogeneous structures in biomembranes.
Raft domains may be involved in certain biocellular functions, such as signal transduction and membrane trafficking~\cite{simons97, simons11}.
In lipid vesicles composed of saturated and unsaturated lipids, the membranes undergo phase separation to afford a solid-ordered (S$_{\rm o}$) phase that is rich in saturated lipids and a liquid-disordered (L$_{\rm d}$) phase that is rich in unsaturated lipids.
In ternary lipid vesicles with cholesterol as the third component, phase separation is observed between L$_{\rm d}$ phase and a liquid-ordered (L$_{\rm o}$) phase rich in saturated lipids and cholesterol.
Phase separation close to the room temperature has been reported for many lipid mixtures. 
The phase-separated structure disappears upon increasing the temperature owing to the mixing entropic contribution.
This influence of temperature on phase separation determines the thermodynamic properties of lipid membranes.
It is also important to control the phase separation under isothermal conditions to understand the characteristics of signal transduction as well as realize industrial applications, such as functional carriers for drug delivery systems~\cite{samad07}.

Biomembranes may contain several types of negatively charged lipids, such as phosphatidylserine, phosphatidylglycerol, phosphatidic acid, and phosphatidylinositol.
These negatively charged lipids play important roles in protein adsorption onto the lipid membranes via electrostatic attraction, the generation of a membrane potential, and channel activity~\cite{mclaughlin95, poyry16, ma17, cabanos17}.
Over the past decade, the phase separation of negatively charged lipid vesicles has been experimentally examined~\cite{shimokawa10, himeno14, kubsch16}.
Phase separation is suppressed in vesicles containing charged unsaturated lipids compared to neutral lipid vesicles because the electrostatic repulsion between the charged unsaturated lipids prevents the formation of a charged lipid-rich phase.
However, phase separation can be enhanced by the addition of a salt owing to weakening of the electrostatic repulsion due to the strong screening effect.
The mechanisms underlying these behaviors have been evaluated using theoretical models based on Poisson-Boltzmann theory and numerical simulations~\cite{may02, shimokawa11, shimokawa16, ito16}.
Therefore, in the case of charged lipid vesicles, salt addition is one way to control phase separation without altering the temperature.

In hypotonic solutions, the phase separation in neutral lipid vesicles can be enhanced by osmotic stress~\cite{hamada11, tik20}.
Since the lipid membranes are semipermeable, vesicles can shrink or swell depending on the osmotic pressure, which is important for metabolism, growth, development, and homeostasis of body fluids.
In addition, membrane lateral tension caused by osmotic pressure also influences the activity of ion channels~\cite{sukharev94, sachs10}.
The osmotic swelling suppresses membrane fluctuation and reduces the fluctuation entropy.
The shape fluctuation in phase-separated membranes is smaller than that in homogeneous membranes owing to the presence of the rigid ordered phases.
Therefore, the contribution of configuration entropy to the membrane shape fluctuation under osmotic stress becomes small compared with that in the absence of osmotic stress.
Consequently, the interaction energy among lipids becomes dominant and phase separation occurs.
Similar phase separation has been observed in vesicles adsorbed onto solid substrates~\cite{gordon08}.
The phase separation can also be controlled by membrane lateral tension without altering the temperature.
However, it remains unclear whether the phase separation in charged lipid membranes can be induced by osmotic stress.

In this study, we investigated the phase separation of binary charged lipid membranes in hypotonic solutions using fluorescence and confocal laser scanning microscopy.
The studied binary lipid vesicles were composed of the negatively charged unsaturated lipid dioleoylphosphatidylserine (DOPS) and the zwitterionic saturated lipid dipalmitoylphosphatidylcholine (DPPC).
To investigate the effects of the headgroup ionization of DOPS on the phase behavior, we examined the phase separation in hypotonic solutions at different pH values.
Although this system was composed of only two components, namely, DPPC and DOPS, it can be regarded as a quasi-ternary system by taking into account the ionization of DOPS.
As the formed phases contained different fractions of charged lipids, they could be identified experimentally by measuring the degree of adsorption of positively charged particles.
Moreover, the stability of the phase-separated structures is discussed on the basis of coarse-grained molecular dynamics simulations.
First, we explain the experimental procedures and the details of the coarse-grained molecular dynamics simulations.
Next, we describe the experimental observations and the results of the numerical simulations.
Finally, we discuss the mechanism of three-phase coexistence and compare our results with those obtained in other relevant studies.

\section{Materials and methods}
\label{matemeth}

\subsection{Materials}
The zwitterionic saturated lipid, 1,2-dipalmitoyl-{\it sn}-glycero-3-phosphocholine (DPPC) and negatively charged unsaturated lipid, 1,2-dioleoyl-{\it sn}-glycero-3-phospho-L-serine (sodium salt) (DOPS) were purchased from Avanti Polar Lipids, Inc.
Rhodamine B 1,2-dihexadecanoyl-sn-glycero-3-phosphoethanolamine, triethylammonium salt (Rho-DHPE) was obtained from Thermo Fisher Scientific.
1,2-Dipalmitoyl-{\it sn}-glycero-3-phosphoethanolamine-N-(7-nitro-2-1,3-benzoxadiazol-4-yl)(ammonium salt)(NBD-PE) was obtained from Avanti Polar Lipids.
Ultrapure water (specific resistance $\geq$ 18 M$\Omega$) was obtained from a Millipore Milli-Q purification system.

\subsection{Preparation of giant unilamellar vesicles}
Giant unilamellar vesicles (GUVs) were prepared by the natural swelling method.
The lipids (DOPS and DPPC) and fluorescent probes (Rho-DHPE and NBD-PE) were dissolved in 2:1 (v/v) chloroform/methanol solution to afford concentrations of 2 mM for each lipid and 0.1 mM for each fluorescent probe.
The two lipid species were mixed in three mixing ratios of DOPS/DPPC=70/30, 50/50, and 30/70 such that the total volume was 20 $\mu$L.
For fluorescence microscopy, Rho-DHPE (2 $\mu$L) was added.
For confocal laser scanning microscopy, Rho-DHPE (2 $\mu$L) and NBD-PE (4 $\mu$L) were added.
The organic solvents were evaporated under a flow of nitrogen gas and the lipids were further dried in a vacuum desiccator for at least 3 h.
The lipid films were then hydrated with 5 $\mu$L of Milli-Q water at 55 $^{\circ}{\rm C}$ for 10 min (prehydration) followed by 195 $\mu$L of glucose solution or pH-adjusted glucose solution (pH = 5, 6, or 8) for at least 3 h at 37 $^{\circ}{\rm C}$. 
We prepared the desired pH value of HCl or NaOH aqueous solution by measuring pH with a pH meter (LAQUA act D-72, HORIBA, Japan).
Glucose was dissolved in these solutions to prepare a pH-adjusted glucose solution.
The glucose concentration was 100 or 200 mM.
Similarly, we prepared a glucose solution with Milli-Q water without HCl and NaOH.
The measured pH value of this solution was 6.96 $\pm$ 0.14.
Therefore, we refer to the results using glucose solution with Milli-Q water as the results for pH = 7.

\subsection{Microscopic observations under osmotic stress}
Prior to the microscopic observations, water with the same pH as the glucose solution used in the natural swelling method was added to the GUV solution to dilute the glucose concentration of the aqueous solution outside the GUVs, thereby generating a concentration gradient between the inside and outside of the GUVs.
Since the lipid membrane is semipermeable, this applies an osmotic pressure and induces water inflow through the membrane.
In the present study, we used hypotonic solutions with lower concentration than in the vesicles.
Water enters from the outside to the inside of the vesicle as a result of osmotic pressure.
After diluting the GUV solution, the solution was put into 3 min and we observed the GUVs by microscopy.
Here, water was added such that the concentration difference across the lipid membrane $\Delta c$ was 0, 20, 40, 60, 80, or 100 mM.
We used 200 mM glucose solution for $\Delta c=$ 0, 80, and 100 mM. In addition, 100 mM glucose solution was used for $\Delta c=$ 20, 40, and 60 mM.

The GUV solution was placed on a glass coverslip, which was then covered with another smaller coverslip at a spacing of $\it ca.$ 0.1 mm.
The GUVs were then examined using a fluorescence microscope (IX71, Olympus) and a confocal laser scanning microscope (FV-1000, Olympus) at room temperature ($\sim$22.5 $\pm$ 2.5 $^{\circ}{\rm C}$).
To avoid photo-induced oxidation during the observations, we limited the observation time to 90 s and disregarded domains that appeared after excitation light irradiation.
To obtain statistical data, 90 GUVs with diameters of 5 - 20 $\mu$m were examined for each condition (lipid composition and $\Delta c$).
Some GUVs ruptured due to osmotic stress, and the rupture occurs within seconds after applying osmotic pressure as shown in Fig.~S1.
We ignored the ruptured vesicles since the observation was performed after waiting for 3 min and applying the osmotic pressure.

We prepared an observation chamber with two compartments separated by a polycarbonate membrane filter (pore size 0.4 $\mu$m, Whatman)~\cite{hamada09} to be able to monitor the dynamics of phase separation induced by the addition of Milli-Q water.
The lipid film was hydrated by 200 mM glucose solution to prepare the vesicle solution, since this experiment was performed only at $\Delta c=$100 mM.
The vesicle solution (5 $\mu$L) was placed in the lower compartment, and put the observation chamber on an inverted confocal laser scanning microscope (FV-1000, Olympus).
After focusing on the sample, we added Milli-Q water (5 $\mu$L) to the upper compartment.
Milli-Q water in the upper compartment penetrated into the lower compartment through the membrane filter.
The vesicle outer solution diluted and the osmotic pressure causes the vesicles to swell.
$\Delta c$ became gradually larger and finally reached $\Delta c=$100 mM.
The time Milli-Q water was placed in the upper compartment was set to 0 sec.

\subsection{Adsorption of particles onto GUVs}
To identify the different phases generated upon phase separation, we measured the adsorption of positively charged particles onto the GUVs.
Amine-modified polystyrene latex beads (fluorescent orange aqueous suspension, 2.5 wt\%, mean particle size 1 $\mu$m) were purchased from Merck.
The colloidal suspension was diluted with Milli-Q water by a factor of 370.
The diluted colloidal suspension and vesicle solution with $\Delta c$ = 100 mM were mixed in a ratio of 2:35 and the samples were examined using a fluorescence microscope or confocal laser scanning microscope.
We counted the number of particles adsorbed onto the various phases.
The total number of counted particles in all phases was 36.

\subsection{Coarse-grained molecular dynamics simulations}
In our previous coarse-grained molecular dynamics simulations~\cite{ito16,shimokawa19}, which were developed from the model originally proposed by Cooke {\it et al.}~\cite{cooke05}, we simulated the phase separation in binary charged lipid membranes.
In the present study, we extended this model to simulate the phase separation in ternary lipid mixtures.

In this approach, a lipid molecule consists of one hydrophilic head bead and two hydrophobic tail beads, and these beads are connected linearly through springs.
The excluded-volume interaction between two beads separated by a distance $r$ is expressed as a short-range repulsive potential
\begin{equation}
\label{rep}
V_{\rm{re}}(r;b_{i})=\begin{cases}
                  4v\left[ \left( \frac{b_{i}}{r} \right)^{12} - \left( \frac{b_{i}}{r} \right)^{6} + \frac{1}{4} \right], & r \leq r_{\rm{c}}, \\
                  0, & r>r_{\rm{c}},
                 \end{cases}
\end{equation}
where $r_{\rm{c}}=2^{1/6}b_{i}$ is the cut-off length for the excluded-volume interaction, $v=k_{\rm B}T$ is the unit of energy, where $k_{\rm B}$ is the Boltzmann constant and $T$ is the absolute temperature, and $i$ may be head-head, head-tail, or tail-tail.
$b_{\rm{head-head}}=b_{\rm{head-tail}}=0.95 \sigma$ and $b_{\rm{tail-tail}}=\sigma$ were used for stable bilayer formation.
Here, $\sigma$ is a unit of length corresponding to the cross sectional diameter of a single lipid molecule and it was approximately 7\AA.
The stretching and bending potentials of the springs between connected beads are expressed as
\begin{equation}
\label{bond}
V_{\rm{st}}(r)=\frac{1}{2}k_{\rm{st}}(r-\sigma)^{2}
\end{equation}
and
\begin{equation}
\label{bend}
V_{\rm{be}}(\theta)=\frac{1}{2}k_{\rm{be}}(1-\cos \theta)^{2},
\end{equation}
where $k_{\rm{st}}=500v$ and $k_{\rm{be}}=60v$ are the stretching strength of connected beads and the bending stiffness of a lipid molecule, respectively, and
$\theta$ is the angle between adjacent bond vectors.
The hydrophobic attractive interaction between hydrophobic beads is expressed as
\begin{equation}
\label{attr}
V_{\rm{at}}(r)=\begin{cases}
                -v, & r<r_{\rm{c}}, \\
                -v \cos^{2} \left[\frac{\pi(r-r_{\rm{c}})}{2 w_{\rm{c}}} \right], & r_{\rm{c}} \leq r \leq r_{\rm{c}}+w_{\rm{c}}, \\
                0, & r>r_{\rm{c}}+w_{\rm{c}},
                 \end{cases}
\end{equation}
where $w_{\rm{c}}$ is the cut-off length for the attractive potential.
By changing the value of $w_{\rm c}$, the S$_{\rm o}$ and L$_{\rm d}$ phases can be expressed qualitatively; the lipid can be regarded as a saturated lipid (S$_{\rm o}$ phase) when $w_{\rm{c}}$ is large, whereas it can be regarded as an unsaturated lipid (L$_{\rm d}$ phase) when $w_{\rm{c}}$ is small~\cite{cooke05}. 
To represent the charged lipids, we considered the electrostatic interaction among charged head groups~\cite{ito16,shimokawa19}. 
The electrostatic repulsion is described by the Debye-H\"{u}ckel potential
\begin{equation}
\label{elec}
V_{\rm{el}}(r)=v \ell_{\rm{B}} z_{1}z_{2}\frac{\exp(-r/\ell_{\rm{D}})}{r},
\end{equation}
where $\ell_{\rm{B}}=\sigma$ ($\simeq$ 7 \AA) is the Bjerrum length, $z_{1}$ and $z_{2}$ are the valences of the interacting charged headgroups, and $\ell_{\rm{D}}=\sigma \sqrt{\varepsilon k_{\rm{B}}T/n_{0}e^{2}}$ is the Debye screening length, which is related to the bulk salt concentration $n_{0}$.
$\varepsilon$ and $e$ are the dielectric constant of the solution and elementary charge, respectively.
$n_{0}$ was set to $100\,\rm{mM}$.
In the experiments, no salt was used for charge screening.
Although this approach may not be suitable for a quantitative comparison, it is nonetheless considered appropriate for the qualitative behaviors resulting from the long-range repulsive force.
Snapshots for $n_{0}=$ 10 and 1 mM are presented in Fig.~S2.
Because the charged headgroup, phosphatidylserine (PS), possesses a monovalent negative charge, we set $z_{1}=z_{2}=-1$. 
We did not set a cut-off for this screened electrostatic interaction.

Each bead position $\bm{r}_{i}$ obeys the stochastic dynamics described by the Langevin equation
\begin{equation}
\label{langevin}
m\frac{d^{2}\bm{r}_{i}}{dt^{2}}=-\eta\frac{d\bm{r}_{i}}{dt}+\bm{f}^{V}_{i}+\bm{\xi}_{i},
\end{equation}
where $m=1$ and $\eta=1$ are the mass and drag coefficients, respectively, and the index $i$ denotes the $i$-th bead.
The force $\bm{f}^{V}_{i}$ is calculated from the derivatives of the interaction potentials Eqs.(\ref{rep})--(\ref{elec}). 
As the units for the timescale, the constant $\tau=\eta\sigma^{2}/v$ was selected, and the time increment was set to $dt=7.5 \times 10^{-3}\tau$. 
The Brownian force $\bm{\xi}_{i}$ satisfies the fluctuation-dissipation theorem
\begin{equation}
\label{flu-dis}
\left< \bm{\xi}_{i}(t)\bm{\xi}_{j}(t')\right>=6v\eta\delta_{ij}\delta(t-t').
\end{equation}
where $\delta_{ij}$ is the Kronecker delta and $\delta(t-t')$ is the Dirac delta function.

We simulated a ternary mixture comprising a negatively charged unsaturated lipids (A-lipid), a neutral unsaturated lipids (B-lipid), and a neutral saturated lipid (C-lipid), where the calculated composition (in terms of number of lipids) was A/B/C = 1000/1000/3000.
The A-, B-, and C-lipids correspond to charged DOPS (denoted DOPS(-)), protonated DOPS (denoted DOPS(N)), and DPPC, respectively.
The initial state of the calculation was a spherical lipid bilayer with a homogeneously mixed lipid distribution, as depicted in Fig.~S3a.
The total calculation time was set to $t=0.5\times 7500\tau$, during which the phase separation dynamics adequately relaxed.
The calculations were performed six times to ensure data reproducibility and obtain representative results.

\section{Results and Discussion}
\label{result}
\subsection{Phase separation induced by osmotic stress}
\begin{figure}[t!]
\begin{center}
\includegraphics[scale=0.9]{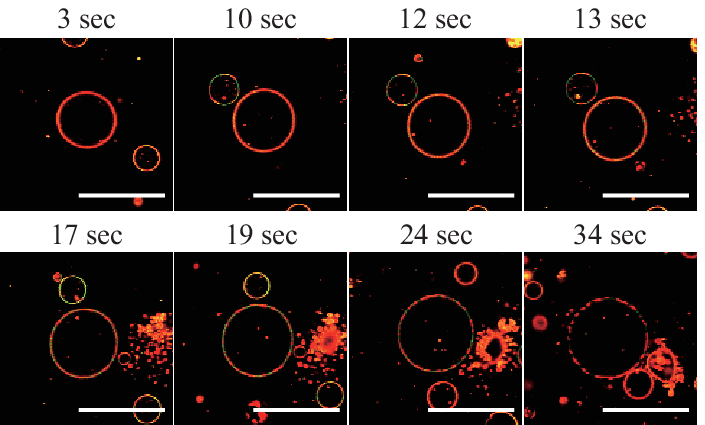}
\end{center}
\caption{\textsf{
Typical phase separation dynamics from homogeneous phase to phase-separated state induced by the addition of Milli-Q water.
The vesicles are composed of DOPS/DPPC=30/70 at pH=7.
The glucose concentration difference $\Delta c$ will reach 100 mM when the vesicle solution and added Milli-Q water are completely mixed each other.
The time Milli-Q water was added to the observation chamber was set 0 sec.
Red and green colors represent Rho-DHPE and NBD-PE, respectively.
Scale bars are 10 $\mu$m.
}}
\label{fig1}
\end{figure}

In order to clarify whether osmotic pressure induces phase separation, Milli-Q water was added to the vesicles with the observation chamber, and the phase separation dynamics was observed in real time as shown in Figure~\ref{fig1}.
The vesicles are composed of DOPS/DPPC=30/70 at pH = 7.
The glucose concentration difference across the lipid membrane, $\Delta c$, reached 100 mM finally, when the solutions in the upper and lower compartments were completely mixed with each other.
Just after Milli-Q water was placed on the chamber, we observed a homogeneous vesicle at 3 sec.
Around 12 sec, the inhomogeneous fluorescent distribution was found and the phase separation took place.
The inhomogeneity of fluorescent probes grew over time and we can see the clear phase separation at 34 sec.
Therefore, phase separation can be observed immediately after Milli-Q water was placed on the chamber, and it finished within tens of second.

\begin{figure*}[t!]
\begin{center}
\includegraphics[scale=0.9]{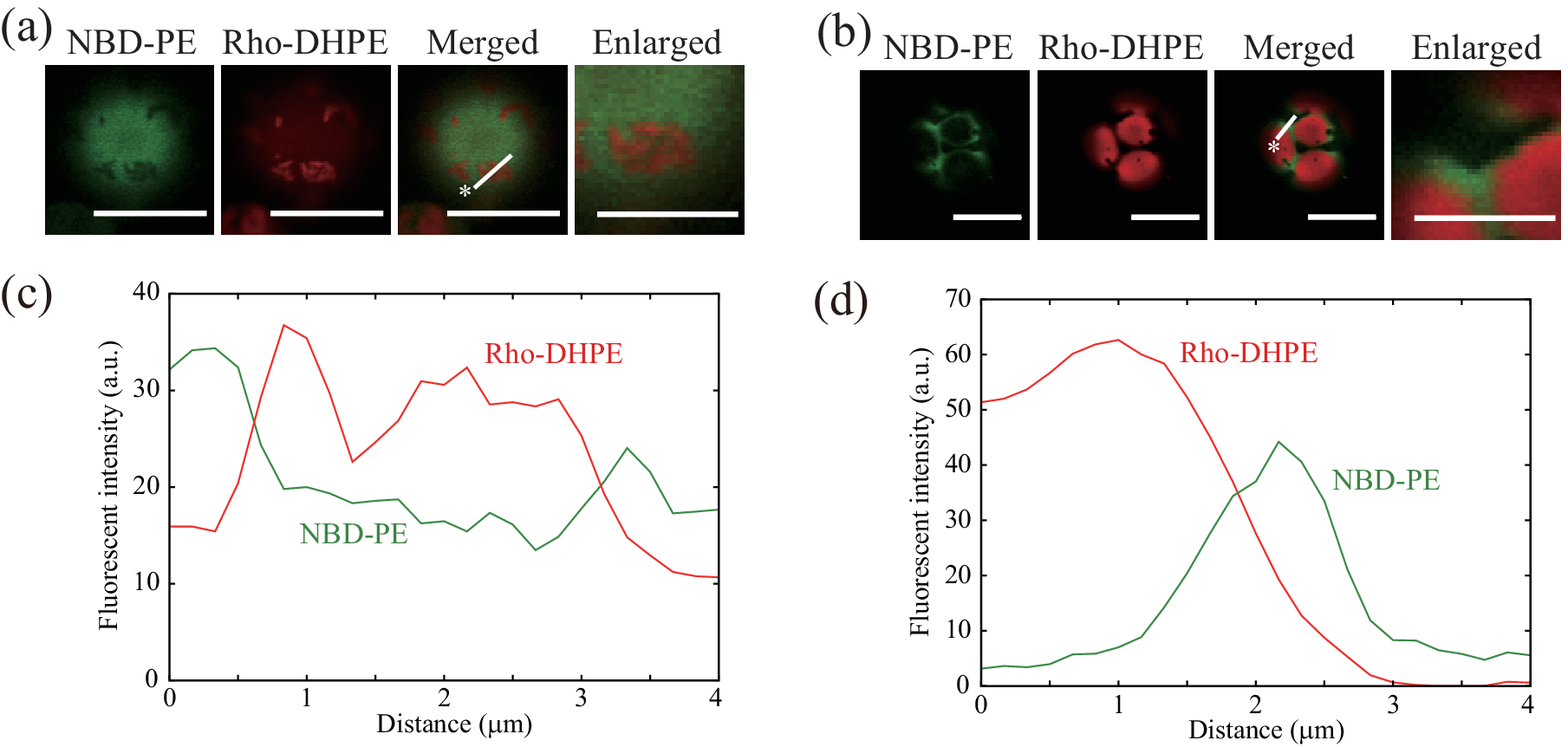}
\end{center}
\caption{\textsf{
Microscopic observation results for pH values of 7.
Confocal laser scanning microscopy images for DOPS/DPPC = 30/70 with $\Delta c=$ 100 mM representing (a) two-phase coexistence and (b) three-phase coexistence.
Images obtained with Rho-DHPE, NBD-PE, and merging Rho-DHPE and NBD-PE and the enlarged view near the white line in the merged image.
Scale bars in Rho-DHPE, NBD-PE, and the merged images are 10 $\mu$m and those in the enlarged images are 5 $\mu$m.
(c,d) Fluorescence intensity profiles along the white lines from the edge indicated with an asterisk in the merged images in panel (a) and (b), respectively.
The green and red lines represent the fluorescence intensity profiles for NBD-PE and Rho-DHPE, respectively.
}}
\label{fig2}
\end{figure*}

We observed the phase separation at $\Delta c=$ 100 mM for DOPS/DPPC = 70/30, 50/50, and 30/70 at pH = 7.
When $\Delta c=0$ mM, almost no phase separation was observed in all examined lipid compositions and this result is consistent with previous study~\cite{shimokawa10}.
On the other hand, we found the phase separation of vesicles in a hypotonic solution, and the typical phase-separated structure observed by confocal laser scanning microscopy is presented in Figure~\ref{fig2}a.
Since Rho-DHPE localized in the unsaturated-lipid-rich L$_{\rm d}$ phase and NBD-PE localized in a relatively ordered phase, the green region corresponds to the DPPC-rich S$_{\rm o}$ phase.
The fluorescence intensity profiles along the white line in the merged image of Figure~\ref{fig2}a are plotted in Figure~\ref{fig2}c.
The two phases appear: an NBD-PE-rich green region (DPPC-rich phase) with $f_{\rm NBD}>f_{\rm Rho}$ (distance from the asterisk: 0 - 0.6 and 3.1 - 4 $\mu$m in Figure~\ref{fig2}c) and a Rho-DHPE-rich red region (DOPS-rich phase) with $f_{\rm NBD}<f_{\rm Rho}$ (0.6 - 3.1 $\mu$m), where $f_{\rm NBD}$ and $f_{\rm Rho}$ indicate the fluorescent intensities of NBD-PE and Rho-DHPE, respectively.

Interestingly, the three-phase coexistence was observed especially in the case of DOPS/DPPC = 30/70, although the lipid membranes consisted of only two components (DOPS and DPPC).
The three-phase coexistence observed by confocal laser scanning microscopy is presented in Figure~\ref{fig2}b.
A dark region that did not contain any fluorescent probes was observed besides the red and green regions.
From the fact that the fluorescent probes were hardly partitioned into the well-ordered phase, we considered that the dark region was the DPPC-rich S$_{\rm o}$ phase.
The details of red and green regions will be discussed later.
To clearly visualize the three-phase coexistence, the fluorescence intensity profiles along the white line in the merged image of Figure~\ref{fig2}b are plotted in Figure~\ref{fig2}d.
The three phases are clearly observed: a Rho-DHPE-rich red region with $f_{\rm NBD}<f_{\rm Rho}$ (distance from the asterisk: 0 - 1.5 $\mu$m in Figure~\ref{fig2}d), an NBD-PE-rich green region with $f_{\rm NBD}>f_{\rm Rho}$ (1.5 - 3 $\mu$m), and a dark region (DPPC-rich phase) with $f_{\rm NBD} \simeq f_{\rm Rho} \simeq 0$ (3 - 4 $\mu$m).
The region where both Rho-DHPE and NBD-PE fluorescence intensities are low appeared only in three-phase coexistence, and this is the clear difference between two- and three-phase coexistences.
The fluorescence intensities for Rho-DHPE and NBD-PE gradually changed at the phase boundary between the red and green regions, as observed in the range between 1 and 2 $\mu$m in Figure~\ref{fig2}d.
Furthermore, the fluorescence intensity for NBD-PE suddenly decreased at the phase boundary between the green and black regions in the range between 2.5 and 3 $\mu$m in Figure~\ref{fig2}d.
These observations imply that the red and green regions were weakly separated with a small line tension at the phase boundary, whereas the green and dark regions were clearly separated with a large line tension.
Therefore, the red and green regions were predominantly composed of DOPS but can be considered to possess distinct properties.
We consider that there were two types of DOPS with different ionization states, namely, negatively charged DOPS (DOPS(-)) and nondissociated (neutral) DOPS (DOPS(N)).
Typical microscopy images of the three-phase coexistence observed for DOPS/DPPC = 30/70 at $\Delta c=$ 20 and 60 mM are presented in Fig.~S4.

To confirm the validity of the obtained results, we additionally examined the effects of the glucose concentration, type of sugar, and the nature of the fluorescent probes.
We observed the phase separation for various glucose concentrations $c$ while the concentration difference was maintained at $\Delta c=0$ mM, as shown in Fig.~S5.
The phase separation did not substantially change as the glucose concentration was varied.
Therefore, the phase separation was induced by the glucose concentration difference across the lipid membranes, while the presence of glucose itself did not affect the phase behavior.
We also observed the phase separation using sucrose instead of glucose.
As shown in Fig.~S6, the three-phase coexistence can be found for DOPS/DPPC=30/70 at $\Delta c=$ 100 mM.
This result indicated that the osmotic pressure-induced phase separation did not depend on the type of sugar.
Furthermore, we monitored the phase separation using additional fluorescent probes and measured the fluorescence intensities as plotted in Fig.~S7.
The three-phase coexistence was observed even using these fluorescent probes.
Thus, we consider that the three-phase coexistence is independent of the choice of fluorescent probe.

\subsection{Identification of three phases}
In Figure~\ref{fig1}b, we show the three phases, namely, the red, green, and dark regions.
As we discussed above, the dark region is the DPPC-rich S$_{\rm o}$ phase.
However, it is not yet known whether the red and green regions are composed of DOPS(-) or DOPS(N).
To elucidate the identities of these two phases, we investigated the main components in the red and green regions by performing experiments using positively charged particles and conducting coarse-grained molecular dynamics simulations.
First, we added positively charged particles from outside the GUVs.
Positively charged particles would be expected to selectively adsorb onto the DOPS(-)-rich phase through electrostatic attraction.
As the addition of positively charged particles to negatively charged GUVs could exert pronounced effects on the phase behavior, these experiments were performed at low particle concentrations where the three-phase coexistence was stably observed.

\begin{figure}[t!]
\begin{center}
\includegraphics[scale=1]{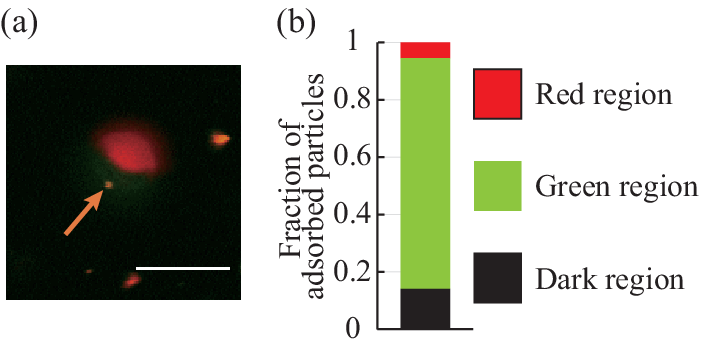}
\end{center}
\caption{\textsf{
(a) Confocal laser scanning microscopy image of phase-separated vesicles with adsorbed positively charged particles.
The adsorbed particle is indicated by the arrow.
Scale bars are 10 $\mu$m.
(b) Fraction of adsorbed particles on each phase-separated region.
The color of the bars (red, green, and black) correspond to the color of each region in the phase-separated images obtained from confocal laser scanning microscopy (Figure~\ref{fig2}b and Figure~\ref{fig3}a).
The total number of counted particles in all phases was 36.
}}
\label{fig3}
\end{figure}

The obtained confocal laser scanning microscopy image is presented in Figure~\ref{fig3}a.
It can be seen that the particles adsorbed onto the green region.
We counted a total of 36 particles and the result is summarized in Figure~\ref{fig3}b.
Approximately 80\% of which (29 particles) adsorbed onto the green region, despite the fact that the area of this region was relatively small compared to the other regions.
For neutral phase-separated membranes with electrically neutral nano/microparticles, Hamada {\it et al.} reported that particles larger than 200 nm tended to adhere to a disordered phase, whereas smaller particles tended to adhere to an ordered phase~\cite{hamada12}.
As we used particles with a size of 1 $\mu$m, they preferentially adsorbed onto the L$_{\rm d}$ phase rather than the S$_{\rm o}$ phase.
Although the particles should be adsorbed to the same extent in the green and red regions composed of DOPS in terms of membrane phase, particles selectively adsorbed onto green region.
Since the electrostatic attraction between the negatively charged region and positively charged particles is expected to be dominant, we consider that the green region mainly consisted of DOPS(-).
Consequently, the red region in Figure~\ref{fig3}a was composed of DOPS(N).

We should note the relationship between the coexisting phases and fluorescent probes.
NBD-PE was localized in the DPPC-rich phase for two-phase coexistence (Figure~\ref{fig2}a), while it was localized in the DOPS(-)-rich phase for three-phase coexistence (Figure~\ref{fig2}b).
Although NBD-PE was partitioned into relatively ordered phases, it cannot be partitioned into well-ordered phases.
Therefore, the DPPC-rich phase in two-phase coexistence may be loosely packed than that in three-phase coexistence.

\subsection{Tension-induced phase separation under different pH conditions}
We consider that the three phases predominantly consisted of DPPC, DOPS(-), and DOPS(N).
If the ionization fraction of DOPS changes, the ratio between DOPS(-) and DOPS(N) also changes, thus affecting the phase behavior.
Therefore, we examined the tension-induced phase separation under various pH conditions.

\begin{figure*}[t!]
\begin{center}
\includegraphics[scale=0.9]{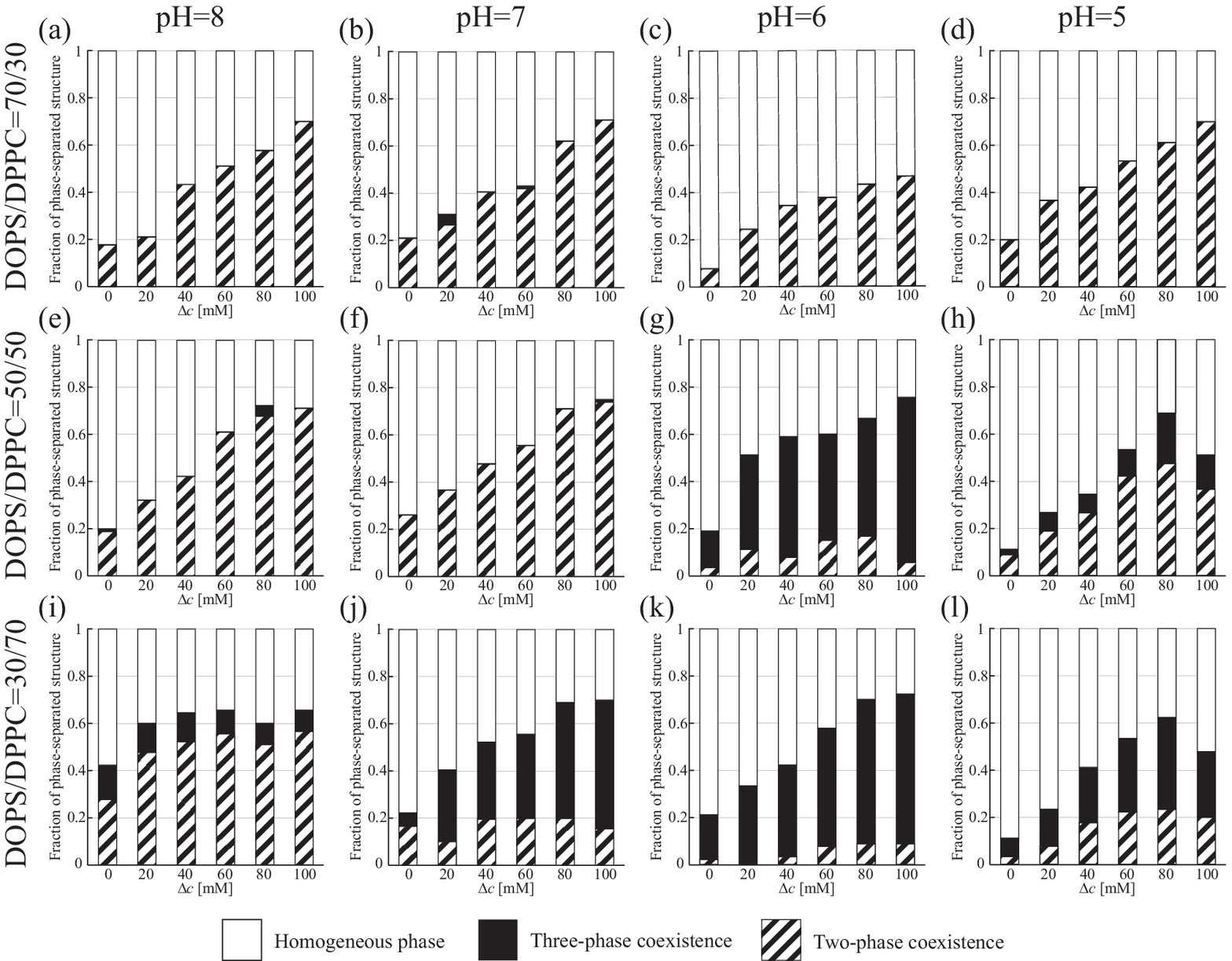}
\end{center}
\caption{\textsf{
Fraction of phase-separated structure as a function of the concentration difference across the lipid membranes $\Delta c$.
The top (a, b, c, d), middle (e, f, g, h), and bottom (i, j, k, l) rows show the results for DOPS/DPPC ratios of 70/30, 50/50, and 30/70, respectively.
The first (a, e, i), second (b, f, j), third (c, g, k), and fourth (d, h, l) columns from left show the results for pH values of 8, 7, 6, and 5, respectively.
The white, black, and hatched bars represent the homogeneous phase, three-phase coexistence, and two-phase coexistence, respectively.
}}
\label{fig4}
\end{figure*}

In Figure~\ref{fig4}, the phase behaviors for solutions possessing different pH values (pH = 8, 7, 6, or 5) are summarized.
Representative microscopy images showing the three-phase coexistence observed at $\Delta c$ = 100 mM  for a DOPS/DPPC ratio of 30/70 and pH values of 8, 6, and 5 are also presented in Fig.~S8.
For all of the compositions and pH values, a tendency toward enhanced phase separation with increasing $\Delta c$ was observed.
This result is qualitatively consistent with previous reports for neutral lipid vesicles~\cite{hamada11, tik20} and the osmotic pressure induced phase separation even for the charged lipid vesicles.

Here, we focus on the case of DOPS/DPPC = 30/70, because the three-phase coexistence was observed for this composition at pH = 7 as shown in Figure~\ref{fig2}b.
The fraction of the three-phase coexistence at pH = 8 was approximately 10\% at all $\Delta c$ values (Figure~\ref{fig4}i).
This increased to 30-50\% at pH = 7 (Figure~\ref{fig4}j) and further increased to 60\% at pH = 6 (Figure~\ref{fig4}k).
However, the fraction of the three-phase coexistence decreased to 20-30\% at pH = 5 (Figure~\ref{fig4}l).
Therefore, the fraction of three-phase coexistence was highest at pH = 6.
In particular, we could observe the three-phase coexistence in the case of DOPS/DPPC = 50/50 at pH = 6 (Figure~\ref{fig4}g), whereas it was not detected at pH = 7 (Figure~\ref{fig4}f).

Therefore, the occurrence of three-phase coexistence was strongly dependent on the solution pH.
This may be attributable to the degree of ionization of the DOPS headgroup, which is determined by the pH.

\subsection{Coarse-grained molecular dynamics simulation}
To clarify which lipid formed circular domains and which lipid formed band-shaped domains surrounding the circular domains upon the occurrence of three-phase coexistence, we further performed coarse-grained molecular dynamics simulations.
On the basis of the experimental results, the three phases predominantly consisted of DPPC, DOPS(N), and DOPS(-).
We thus simulated the phase separation for a ternary lipid vesicle composed of a negatively charged unsaturated lipid (A-lipid) mimicking DOPS(-), a neutral unsaturated lipid (B-lipid) mimicking DOPS(N), and a neutral saturated lipid (C-lipid) mimicking DPPC with a calculated composition of A/B/C = 1000/1000/3000.
Because the attraction between saturated lipids is stronger than that between unsaturated lipids, and the A- and B-lipids had the same unsaturated hydrocarbon tails, we set $w_{\rm c}^{\rm AA}=w_{\rm c}^{\rm BB}<w_{\rm c}^{\rm CC}$.
Furthermore, the A- and B-lipids should be weakly phase separated owing to their identical unsaturated hydrocarbon chains, whereas these two lipids should be strongly phase separated from the C-lipid as a result of the different hydrocarbon tails.
Therefore, we set $w_{\rm c}^{\rm AB}>w_{\rm c}^{\rm BC}=w_{\rm c}^{\rm AC}$.
We investigated the arrangement of the three phases on the surface of a vesicle displaying three-phase coexistence.
Therefore, $w_{\rm c}^{\rm AA}=w_{\rm c}^{\rm BB}>w_{\rm c}^{\rm AB}$ was assumed to result in the stable formation of an A-lipid-rich phase and a B-lipid-rich phase.
Finally, to represent the physical properties of these lipids, we set $w_{\rm c}^{\rm AA}=1.7 \sigma$, $w_{\rm c}^{\rm BB}=1.7 \sigma$, $w_{\rm c}^{\rm CC}=1.8 \sigma$, $w_{\rm c}^{\rm AB}=1.55 \sigma$, $w_{\rm c}^{\rm BC}=1.5 \sigma$, and $w_{\rm c}^{\rm AC}=1.5 \sigma$.

\begin{figure}[t!]
\begin{center}
\includegraphics[scale=1]{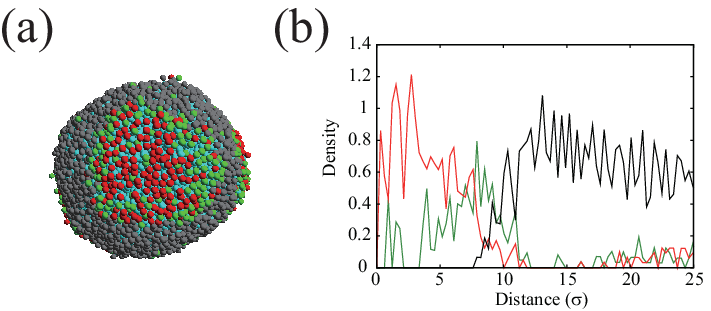}
\end{center}
\caption{\textsf{
(a) Snapshot of a phase-separated vesicle obtained by coarse-grained molecular dynamics simulations.
The green, red, and gray beads represent the negatively charged unsaturated lipids (A-lipids), neutral unsaturated lipids (B-lipids), and neutral saturated lipids (C-lipids), respectively.
(b) Bead density profile from the center of the domain.
The green, red, and black lines represent the bead densities of A-, B-, and C-lipids, respectively.
}}
\label{fig5}
\end{figure}

Figure~\ref{fig5}a shows a representative phase-separated vesicle with the domain in the center.
Snapshots of the phase separation dynamics and the time evolution of the attractive potential and electrostatic interaction are presented in Fig.~S3, where the green, red, and gray beads indicate the negatively charged unsaturated lipids (A-lipids, DOPS(-)), neutral unsaturated lipid (B-lipids, DOPS(N)), and neutral saturated lipids (C-lipids, DPPC), respectively.
We could thus reproduce the three-phase coexistence in coarse-grained molecular dynamics simulations.
The calculations were performed six times to ensure data reproducibility, and we observed three or four domains on the vesicles in all of the calculations.
All of these were circular domains formed by the B-lipids surrounded by the A-lipid-rich phase.
This domain structure is qualitatively identical to the experimental observations shown in Figure~\ref{fig2}b and Figure~\ref{fig3}a.
The calculated lipid density profiles from the center of the domain are plotted in Figure~\ref{fig5}b.
The red B-lipid-rich region was the closest to the center of the domain, followed by the green A-lipid-rich region, while the gray C-lipid-rich region appeared outside of the domain.
These profiles are essentially the same as the fluorescence intensity profiles shown in Figure~\ref{fig2}d.
Therefore, it was demonstrated that the structure in which the DOPS(-)-rich phase is sandwiched between the DOPS(N)-rich and DPPC-rich phases was energetically stable.

\subsection{Discussion}
Similar to how osmotic stress induces phase separation in neutral lipid membranes, it also promotes phase separation in charged lipid membranes.
The mechanism underlying the phase separation induced by osmotic stress was interpreted as the loss of membrane fluctuation entropy~\cite{tik20,gordon08}.
A homogeneous membrane with an L$_{\rm d}$ phase is soft and greatly fluctuates.
In contrast, a phase-separated membrane with an ordered phase fluctuates less than homogeneous membranes.
Thermal membrane fluctuation is suppressed by the membrane tension induced by the osmotic stress.
Therefore, the loss of the membrane fluctuation entropy under osmotic stress for a homogeneous membrane exceeds that for a phase-separated membrane.
Consequently, the homogeneous membrane becomes relatively more unstable than the phase-separated membrane and the phase-separated membrane becomes more stable as the osmotic stress increases.
As phase separation is also induced in negatively charged lipid membranes by osmotic stress, it was found that the effect of osmotic stress overcomes the electrostatic repulsion due to the formation of a charged-lipid-rich phase.

\begin{figure}[t!]
\begin{center}
\includegraphics[scale=1]{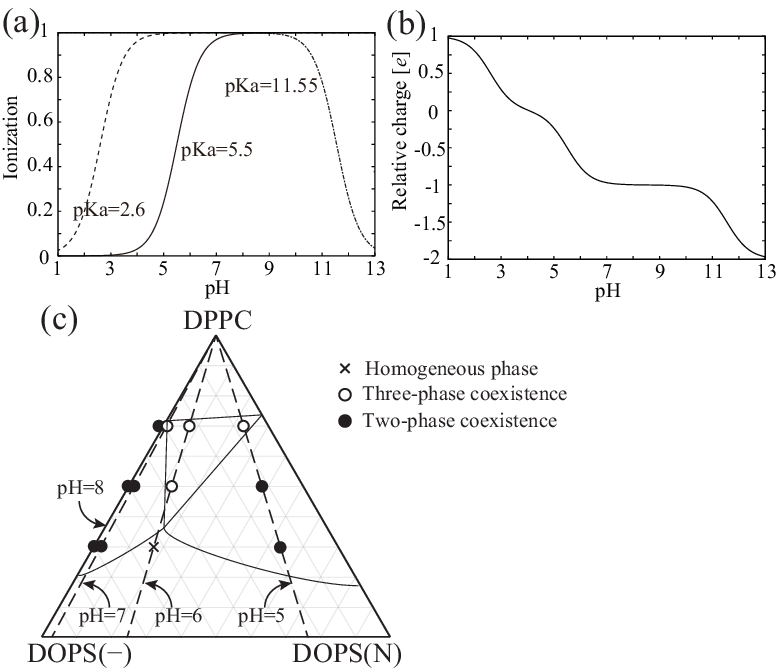}
\end{center}
\caption{\textsf{
(a) Fraction of ionized polar groups in the DOPS headgroup as a function of pH.
The solid, dashed, and dot-dashed lines represent the ionization fractions of the phosphate, carboxy, and amine groups, respectively.
(b) Relative charge of DOPS as a function of pH.
(c) Gibbs triangle for a quasi-ternary lipid mixture composed of DOPS(-), DOPS(N), and DPPC with $\Delta c=$100 mM.
The cross, open circles, and filled circles represent the homogeneous phase, three-phase coexistence, and two-phase coexistence, respectively.
}}
\label{fig6}
\end{figure}

We observed the three-phase coexistence especially in the case of DOPS/DPPC = 30/70 and considered that the ionization state of the DOPS headgroup may play an important role.
The headgroup of DOPS (i.e., PS) contains three polar groups, namely, phosphate, carboxy, and amine moieties.
As the phosphate and carboxy groups have monovalent negative charges and the amine group has a monovalent positive charge, the net charge of PS is typically -1.
However, the ionic dissociation is dependent on the solution pH.
The fraction of dissociation $\alpha$ can be expressed as
\begin{equation}
\alpha=\begin{cases}
\frac{1}{1+10^{{\rm pK_{a}}-{\rm pH}}} ~~~~~\mbox{for the phosphate and carboxy groups}, \\
\frac{1}{1+10^{{\rm pH}-{\rm pK_{a}}}} ~~~~~\mbox{for the amine group.}
\end{cases}
\end{equation}
The fraction of ionic dissociation as a function of pH is plotted in Figure~\ref{fig6}a.
As the phosphate and amine groups have constant pK$_{\rm a}$ values of 2.6 and 11.55, respectively, pH changes in the vicinity of pH = 7 do not affect the ionic dissociation of these groups.
In contrast, the ionic dissociation of the carboxy group is sensitive to pH changes in this region owing to its pK$_{\rm a}$ of 5.5~\cite{cevc81}.
On the basis of the fraction of ionization of PS as shown in Figure~\ref{fig6}a, we calculated the relative charge of PS as a function of pH as plotted in Figure~\ref{fig6}b.
At pH = 7 and 8, the relative charges were estimated to be -0.969 and -0.997, respectively.
This implies that most of the carboxy groups are dissociated, as these values are almost -1.
As the pH decreases, the fraction of dissociated carboxy groups also decreases, and the relative charges changed to -0.759 at pH = 6 and -0.236 at pH = 5.
Therefore, the systems at pH = 7 and 8 are almost binary lipid mixtures consisting of DOPS(-) and DPPC.
In contrast, the systems at pH = 6 and 5 can be regarded as ternary lipid mixtures composed of DOPS(-), DOPS(N), and DPPC.
In the experimental observations for the case of DOPS/DPPC = 30/70, as the pH was decreased from pH = 8, the fraction of three-phase coexistence increased until pH = 6 and then decreased from pH = 6 to 5.
Since the fraction of DOPS(-) decreases while that of DOPS(N) increases until pH = 6, the occurrence of three-phase coexistence is favored.
At pH = 5, the fraction of DOPS(N) becomes larger than that of DOPS(-) and the system approaches a binary mixture consisting solely of DPPC and DOPS(N).
Consequently, the three-phase coexistence is suppressed owing to the reduced amount of DOPS(-).
Finally, we summarize the phase behavior of the quasi-ternary lipid mixture for $\Delta c=$100 mM in the Gibbs triangle presented in Figure~\ref{fig6}c.

Although the actual lipid compositions at pH = 7 and 8 were almost identical, the fractions of three-phase coexistence were different.
Moreover, we observed the three-phase coexistence, although the systems at pH = 7 and 8 were almost binary lipid mixtures.
The value of pK$_{\rm a}$ is dependent on the structure of the molecular assembly and the lipid composition and possibly also on the membrane tension.
Based on our experimental results, it is natural to consider that the pK$_{\rm a}$ may exceed larger than 5.5.
In fact, some previous studies have also indicated that the apparent pK$_{\rm a}$ of PS is greater than 5.5~\cite{bajagur09}.
Therefore, it is important to accurately measure the pK$_{\rm a}$ value by titration analysis in the near future.

\begin{figure}[t!]
\begin{center}
\includegraphics[scale=1]{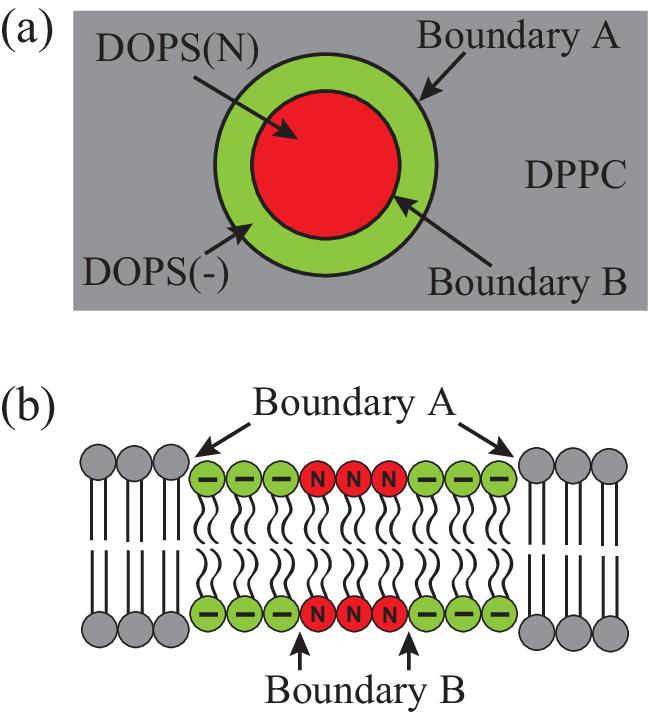}
\end{center}
\caption{\textsf{
(a) Top view and (b) side view of a schematic illustration of the domain structure in the case of three-phase coexistence}}
\label{fig7}
\end{figure}

The experiments involving charged particles and the coarse-grained molecular dynamics simulations revealed that DOPS(N)-rich circular domains successively surrounded by DOPS(-)-rich regions and a DPPC-rich phase were formed.
Here, we discuss the shape of the phase rich in charged lipids (DOPS(-)).
The formation of the charged-lipid-rich phase increases the free energy owing to the electrostatic repulsion between the charged headgroups.
Since the electrostatic interactions are long-range interactions, the total electrostatic interaction is the sum of all interacting charge pairs.
Therefore, a one-dimensional band-shaped domain rather than a two-dimensional circular domain increases the distance between interacting pairs and can thus reduce the total electrostatic interactions.
In other words, as depicted in Figure~\ref{fig7}, the phase rich in charged lipids (DOPS(-)) is in contact with two phases that are poor in charged lipids (DOPS(N) and DPPC) at boundaries A and B, which effectively reduces the electrostatic interactions near the interfaces.
This behavior has already been discussed theoretically~\cite{bossa16}.

Next, we focus on the phase separation between the DOPS(-)- and DOPS(N)-rich phases.
Both phases are rich in unsaturated lipids and are generally mixed with each other.
Furthermore, from the perspective of electrostatic interactions, these two phases should be mixed to reduce the charged lipid concentration and thus the electrostatic interaction.
Nevertheless, phase separation was observed.
Focusing on the interface defined as boundary B in Figure~\ref{fig7}, the line tension is not high as discussed for Figure~\ref{fig2}d.
However, the domain shape fluctuation that indicates lower line tension was not observed experimentally.
Therefore, the physical properties of these two unsaturated lipid-rich phases are significantly different.
We consider that there exists an attractive interaction between DOPS(N) molecules that overcomes the free energy loss due to the electrostatic interactions between DOPS(-) molecules, and that this attraction leads to a significant difference between the DOPS(-)- and DOPS(N)-rich phases.

\begin{figure}[t!]
\begin{center}
\includegraphics[scale=1]{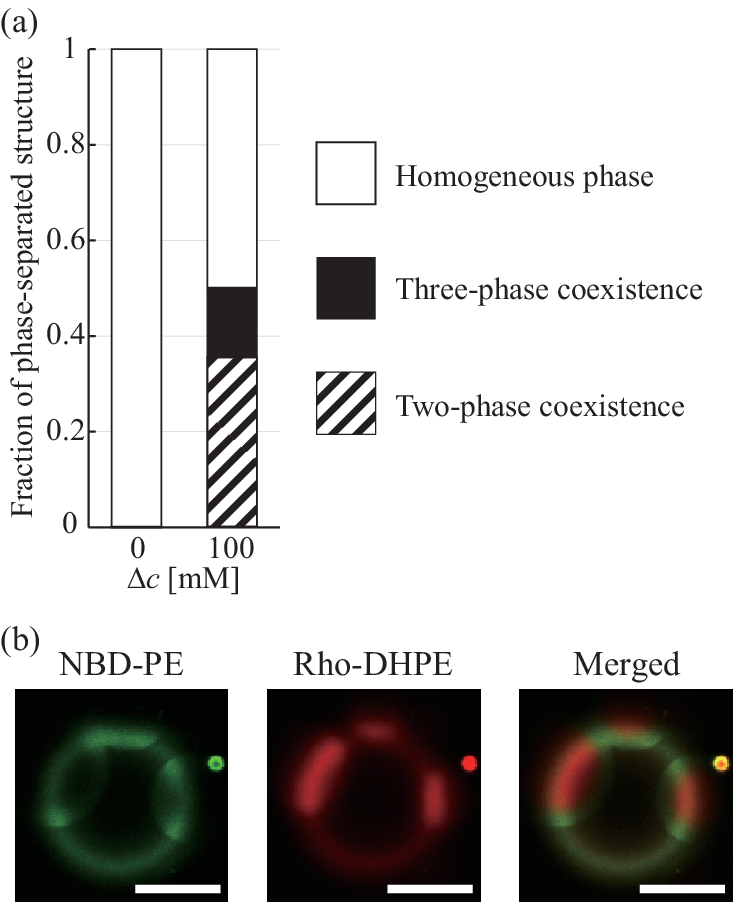}
\end{center}
\caption{\textsf{
(a) Fraction of phase-separated structure as a function of the concentration difference across the lipid membranes $\Delta c$ for DOPS/DPPC = 30/70 in D$_2$O.
The white, black, and hatched bars represent the homogeneous phase, three-phase coexistence, and two-phase coexistence, respectively.
(b) Confocal laser scanning microscopy images for DOPS/DPPC = 30/70 with $\Delta c=$100 mM showing the three-phase coexistence.
Scale bars are 10 $\mu$m.}}
\label{fig8}
\end{figure}

It was reported that the attraction between DOPS(N) molecules may be attributable to intermolecular hydrogen bonding between protonated PS headgroups~\cite{boggs87}.
To test this hypothesis, we conducted similar experiments in D$_2$O instead of Milli-Q water.
As shown in Figure~\ref{fig8}, we observed phase separation in the case of DOPS/DPPC = 30/70 with $\Delta c=$ 100 mM in this solvent system.
Although the phase-separated structures were not detected for $\Delta c=$0 mM, the phase separation increased with increasing $\Delta c$.
These results are essentially the same as those obtained using Milli-Q water, and this phase separation was induced by the solute concentration difference across the membrane.
For $\Delta c$= 100 mM in Milli-Q water, the fraction of three-phase coexistence was approximately 50\% as shown in Figure~\ref{fig4}j.
In D$_2$O, however, this fraction was approximately 10\%, indicating that the occurrence of three-phase coexistence was suppressed.
It is therefore considered that D$_2$O inhibited hydrogen bonding between the protonated DOPS molecules, and thus hindered the three-phase coexistence.
Therefore, the three-phase coexistence was stabilized by the hydrogen bonding between protonated DOPS molecules.
In addition, this attraction was sufficiently strong to form domains even in the case of DOPS, which typically associates loosely owing to its unsaturated hydrocarbon tails.

We also measured fluorescence anisotropy of the fluorescent probe, 1,6-diphenyl-1,3,5-hexatriene (DPH), in DOPS-single-component membranes at various pH values to clarify whether hydrogen bonding between the DOPS(N) headgroups affect the ordering of the hydrocarbon tails of DOPS.
The results are shown in Fig.~S9, and it was found that the fluorescence anisotropy is hardly affected by the change in pH.
In other words, the interactions between hydrocarbon tails are almost identical between DOPS(-) and DOPS(N), that is, DOPS(N)-rich phase is L$_{\rm d}$ phase.
Therefore, this result implies that the attraction between DOPS(N) molecules comes from the hydrophilic parts rather than the hydrophobic parts, which is consistent with the above discussion.

We discussed the ionization of DOPS molecules and the resulting phase separation from pH changes and the adsorption of charged particles.
Since these experiments indirectly indicated the ionization of DOPS, some experiments which measure the local surface charge density directly are desired in the future.
In addition, the local surface charge density strongly influences the localization and regulation of membrane proteins in cell membranes.
For example, Fuhs {\it et al.} used a quantitative surface charge microscopy to measure the local surface charge in the phase-separated supported lipid bilayer membranes~\cite{fuhs18}.
It will be important to directly observe the distributions of DOPS(-) and DOPS(N) in the phase-separated membranes using such a technique.

In Figure~\ref{fig4}h and \ref{fig4}l, the fraction of phase separation was decreased at $\Delta c=$ 100 mM for DOPS/DPPC=50/50 and 30/70.
It was reported that the line tension at phase-separated domain boundary increases with the membrane tension induced by osmotic pressure~\cite{tik20}.
In other words, domain formation from a homogeneous state can cause the large loss of free energy.
Therefore, it was necessary to overcome a large energy barrier for the transition from the homogeneous phase to the most stable phase-separated state (domain nucleation)~\cite{oglecka14}.
If this energy barrier is greater than thermal energy, a metastable homogeneous phase is maintained even for the application of osmotic pressure.
Hence, we speculated that the vesicles cannot be relaxed to the phase-separated state and the fraction of homogeneous phase apparently increases at large $\Delta c$.
In order to reveal this point, it is necessary to measure the line tension at the phase-separated domain boundary in the future.
In particular, the pH dependence of line tension can be an important study.

Bandekar and Sofou reported the phase separation of DOPC/distearylphosphatidylserine (DSPS)/cholesterol while varying the solution pH~\cite{bandekar12}.
Upon the dissociation of DSPS, the protonated DSPS (DSPS(N)) formed domains surrounded by a charged DSPS (DSPS(-))-rich phase, which was ascribed to the attraction between DSPS(N) molecules mediated by hydrogen bonding.
We have obtained fundamentally similar results for different combinations of saturated and unsaturated lipids.
The key differences in our study are that (i) the phase separation was induced by osmotic stress without changing the temperature, (ii) our system is a purely binary system without cholesterol, and (iii) charged unsaturated lipids were used, where the final difference is the most important one.
As DSPS contains long saturated hydrocarbon tails, there exist a strong van der Waals attraction between lipid molecules which is important for forming a domain structure.
In contrast, we used DOPS with unsaturated hydrocarbon tails, for which this attraction is much weaker.
Nevertheless, we observed the phase-separated structure.
Therefore, it is suggested that the attraction via hydrogen bonding between protonated PS moieties is sufficiently strong.
It has thus been demonstrated that domain formation by the ionization state of PS is a universal phenomenon that does not depend on the tail structures of the PS lipids.

Our experimental findings can also be theoretically justified, which will be described in more detail in a subsequent theoretical paper.
Mengistu {\it et al.} have proposed a theoretical model to describe the phase separation in binary charged lipid membranes composed of a negatively charged lipid and a zwitterionic lipid~\cite{mengistu10}.
In particular, this model considered the dependence of the phase behavior on pH and may therefore serve as an important theoretical model to interpret our experimental results.
However, the negatively charged lipid in this theoretical model contained only one negative charge in the headgroup, in contrast to the complex headgroup structure of DOPS containing one positive and two negative charges.
Our experiments have revealed that the intermolecular interaction between DOPS(-) molecules is distinct from that between DOPS(N) molecules owing to the contribution of hydrogen bonding.
Therefore, it is important to propose a model that takes into account such effects in the near future.

\section{Conclusion}
\label{conclu}
We have examined the phase separation caused by osmotic stress in binary charged lipid membranes composed of the charged unsaturated lipid DOPS and zwitterionic saturated lipid DPPC.
The phase separation was induced by the application of osmotic pressure to vesicles, and the coexistence of three phases, namely, those riich in DPPC, charged DOPS (DOPS(-)), and neutral DOPS (DOPS(N)), was observed, especially in the case of DOPS/DPPC = 30/70.
The ionic dissociation of the DOPS headgroup was important for the three-phase coexistence.
Although the studied system was composed of only two components, DPPC and DOPS, it can be regarded as a quasi-ternary system owing to the ionization of DOPS.
Experiments involving the adsorption of charged particles to the generated phases permitted identification of the three phases and revealed that DOPS(N)-rich circular domains were successively surrounded by DOPS(-)-rich and DPPC-rich phases.
Coarse-grained molecular dynamics simulations successfully reproduced this experimentally observed domain arrangement and indicated that the three-phase coexistence was energetically stable.
The experiments involving charged particles and coarse-grained molecular dynamics simulations demonstrated that the domain rich in charged lipids (DOPS(-)) possessed a one-dimensional band shape.
This domain shape increased the distance between the interacting charged lipid pairs, effectively reducing the electrostatic interaction and contributing to the stability of the three coexisting phases.
The three-phase coexistence was stabilized by the hydrogen bonding between DOPS(N).
The details of the interaction between charged lipids, water, and counter ions at nanoscale are still unclear.
It will be important to clarify what interactions in the hydophilic headgroups form ordered structures at microscale using various charged lipids.

Since the body temperature of homeothermic animals does not change significantly, it is important in these organisms that the phase separation of lipid membranes can be modulated under isothermal conditions.
In charged lipid membranes, phase separation can be induced by salt addition.
In addition, in this study, we have showed that the formation of phase-separated structures can be enhanced by osmotic pressure.
The combination of salt addition and osmotic pressure enables more versatile control over phase separation under isothermal conditions.
It is anticipated that the results of this study will prove useful for both attaining a better understanding of the formation of ordered-structures in living organisms and realizing industrial applications, such as functional carriers for drug delivery.

\begin{acknowledgements}
The coarse-grained molecular dynamics simulations were performed using the parallel computer ``SGI UV3000'' at the Research Center for Advanced Computing Infrastructure at JAIST.
We acknowledge support from the Bilateral Joint Research Project (Japan-Slovenia) of the Japan Society for the Promotion of Science (JSPS). 
H.I. acknowledges support from a Grant-in-Aid for Early-Career Scientists (Grant No. JP19K14675) and a Grant-in-Aid for Scientific Research (Grant No. JP19H00749) from JSPS.
Y.H. acknowledges support from a Grant-in-Aid for Scientific Research on Innovative Areas ``Aquatic Functional Materials'' (Grant No. JP19H05718) from JSPS.
N.S. acknowledges support from a Grant-in-Aid for Scientific Research (C) (Grant No. JP17K05610) from JSPS.
M.T. acknowledges support from a Grant-in-Aid for Scientific Research on Innovative Areas ``Thermal Biology'' (Grant No. JP15H05928) from the Ministry of Education, Culture, Sports, Science, and Technology of Japan (MEXT) and a Grant-in-Aid for Scientific Research (B) (Grant No. JP26289311) from JSPS.
\end{acknowledgements}

\bibliography{reference}

\clearpage
\appendix
\setcounter{figure}{0}
\renewcommand{\thefigure}{S\arabic{figure}}
\onecolumngrid
\section*{Supporting information}

\begin{figure}[h!]
\begin{center}
\includegraphics[scale=1]{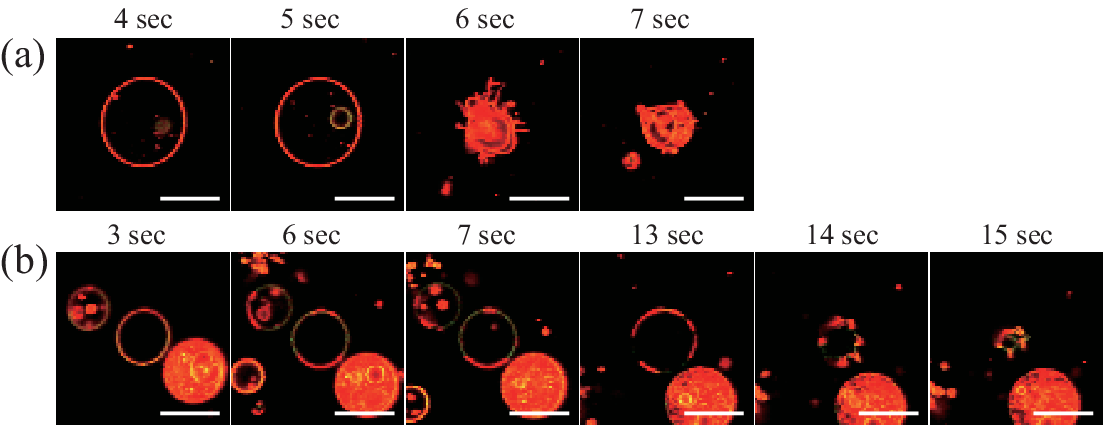}
\end{center}
\caption{
Rupture of homogeneous vesicle and phase-separated vesicle at $\Delta c$ = 100 mM for DOPS/DPPC = 30/70 at pH = 7.
During the observation of phase separation dynamics, some vesicles rupture.
(a) a homogeneous vesicle ruptured at 6 sec.
(b) phase separation occurred at 6 sec and the phase-separated vesicle ruptured at 14 sec.
Scale bars are 10 $\mu$m.
}
\end{figure}

\begin{figure}[h!]
\begin{center}
\includegraphics[scale=0.8]{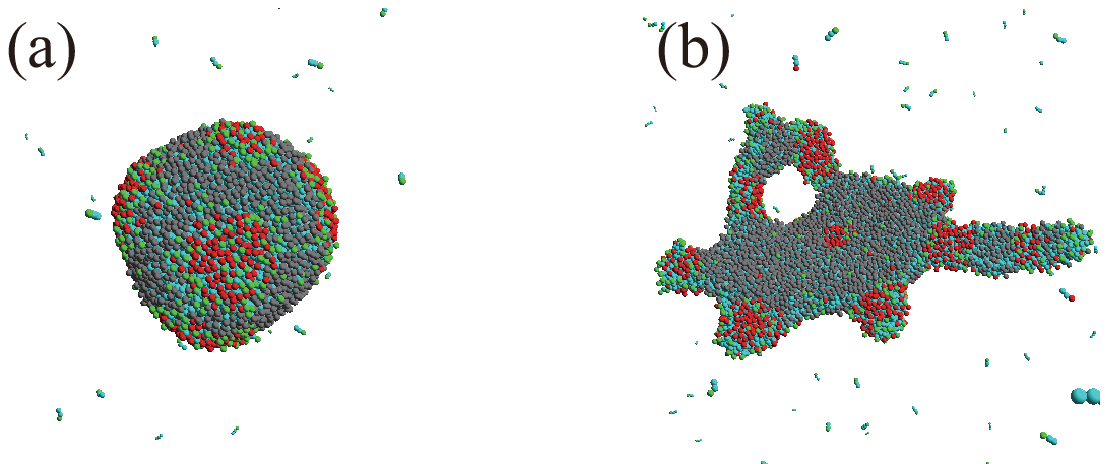}
\end{center}
\caption{
(a) Typical snapshots of a charged vesicle at $t = 0.5 \times 7500 \tau$ for (a) $n_{0}$ = 10 mM and (b) 1mM. The green, red, and gray beads represent the negatively charged unsaturated lipids (A-lipids), neutral unsaturated lipids (B-lipids), and neutral saturated lipids (C-lipids), respectively. Although we observed a phase-separated spherical vesicle for $n_{0}$ = 10 mM, numerous negatively charged unsaturated lipids (A-lipids) had escaped from the vesicle as shown in (a). In contrast, we did not observe any spherical vesicles for $n_{0}$ = 1 mM. Since our focus was to investigate the phase separation on the vesicle surface, we used $n_{0}$ = 100 mM in the main text to observe the phase separation in a stable manner.}
\end{figure}

\begin{figure}[th]
\begin{center}
\includegraphics[scale=0.6]{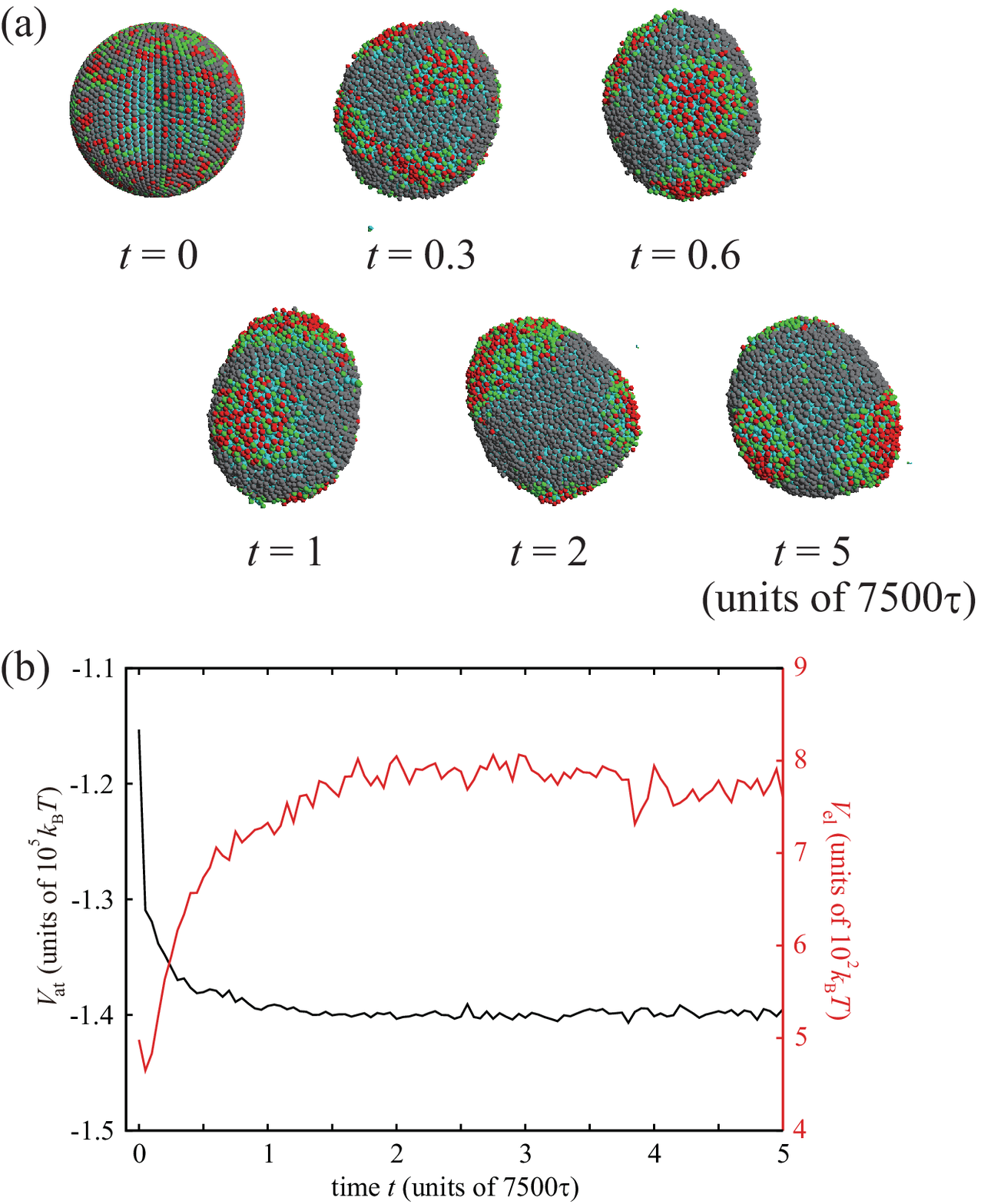}
\end{center}
\caption{
(a) Typical snapshots of the phase separation in a charged vesicle. The green, red, and gray beads represent the negatively charged unsaturated lipids (A-lipids), neutral unsaturated lipids (B-lipids), and neutral saturated lipids (C-lipids), respectively. (b) Time evolution of the attractive potential $V_{\rm at}$ and the electrostatic interaction $V_{\rm el}$, which are represented by black and red lines, respectively.}
\end{figure}

\begin{figure}[t!]
\begin{center}
\includegraphics[scale=0.9]{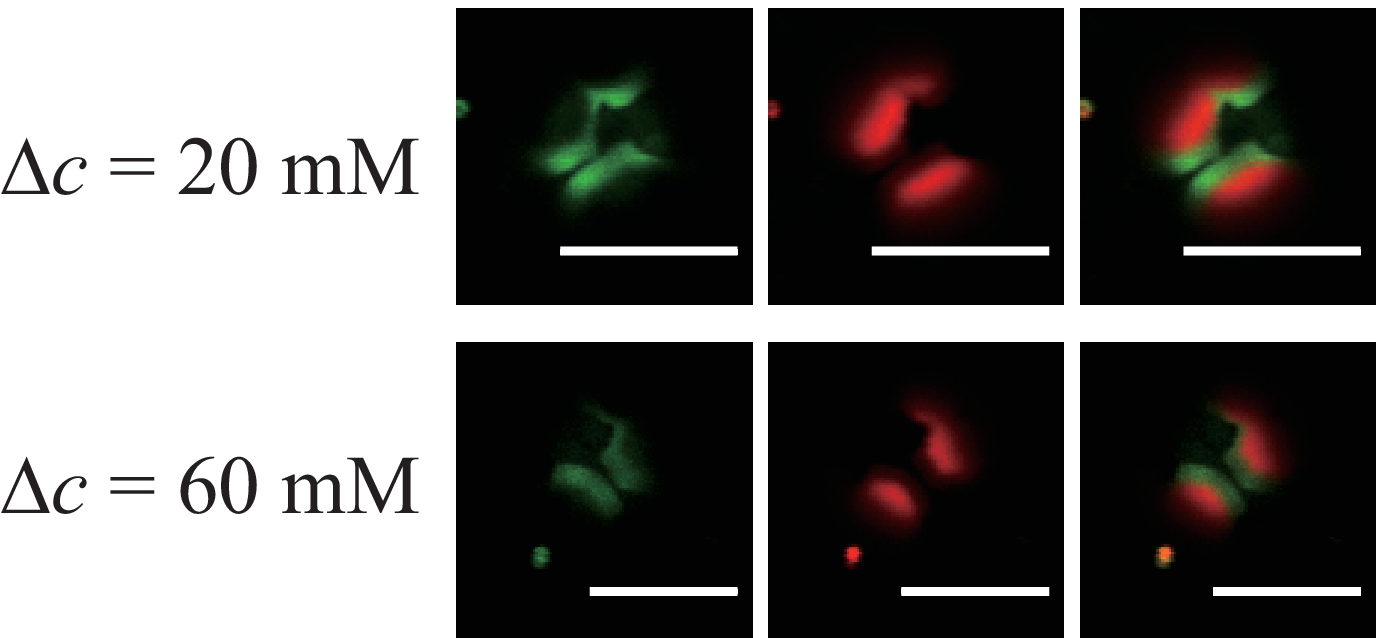}
\end{center}
\caption{
Three-phase coexistence at $\Delta c$ = 20 and 60 mM for DOPS/DPPC = 30/70 at pH = 7. The green and red regions represent NBD-PE-rich and Rho-DHPE-rich regions, respectively. Scale bars are 10 $\mu$m.}
\end{figure}

\begin{figure}[th]
\begin{center}
\includegraphics[scale=0.5]{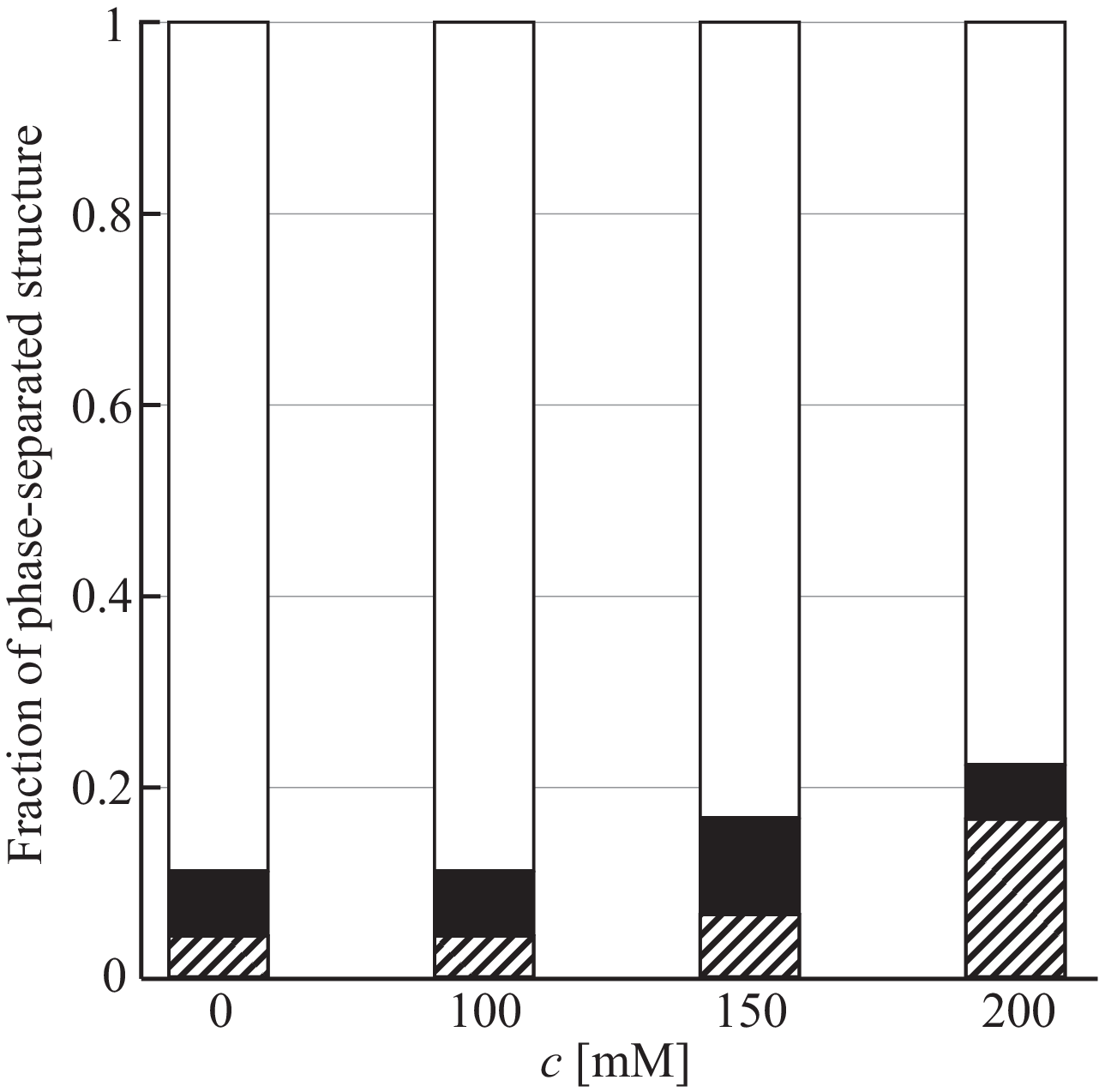}
\end{center}
\caption{
Fraction of phase separated structure for DOPS/DPPC = 30/70 at pH = 7. The vesicles were prepared using glucose solutions with concentration $c$ and the glucose concentration difference $\Delta c$ was fixed at 0. The white, black, and hatched bars represent the homogeneous phase, three-phase coexistence, and two-phase coexistence, respectively. The phase behavior was not influenced by the glucose concentration $c$ and strong phase separation was not induced. Therefore, the phase separation described in the main text was induced by the glucose concentration difference across the lipid membrane $\Delta c$.}
\end{figure}

\begin{figure}[th]
\begin{center}
\includegraphics[scale=0.9]{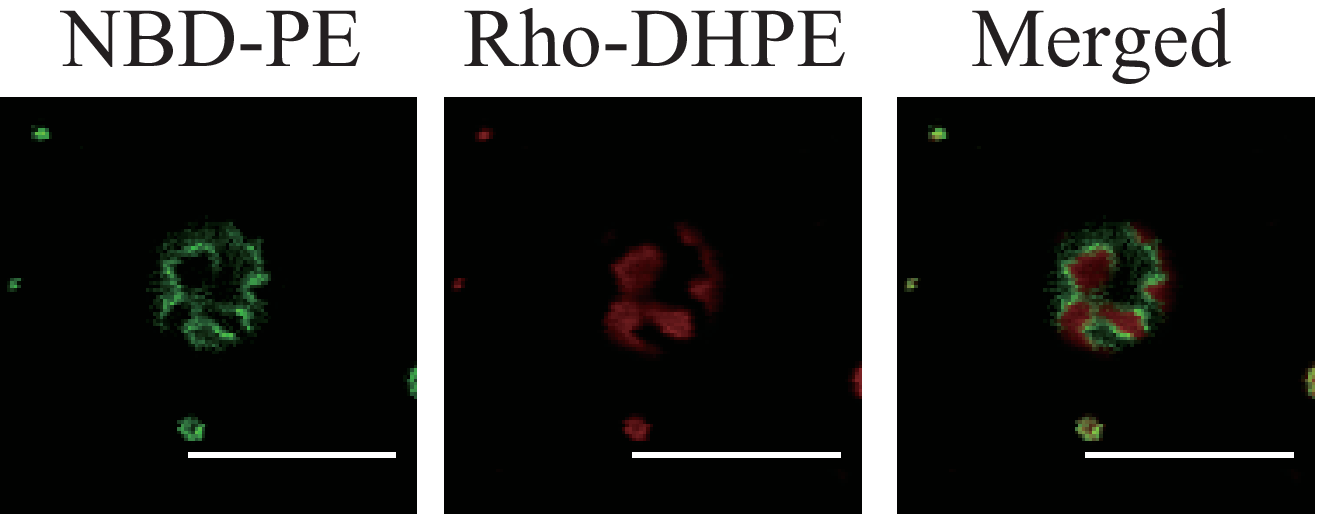}
\end{center}
\caption{
Three-phase coexistence at $\Delta c$ = 100 mM for DOPS/DPPC = 30/70 at pH = 7 using sucrose instead of glucose. The green and red regions represent NBD-PE-rich and Rho-DHPE-rich regions, respectively. Scale bars are 10 $\mu$m.
}
\end{figure}

\begin{figure}[t!]
\begin{center}
\includegraphics[scale=0.7]{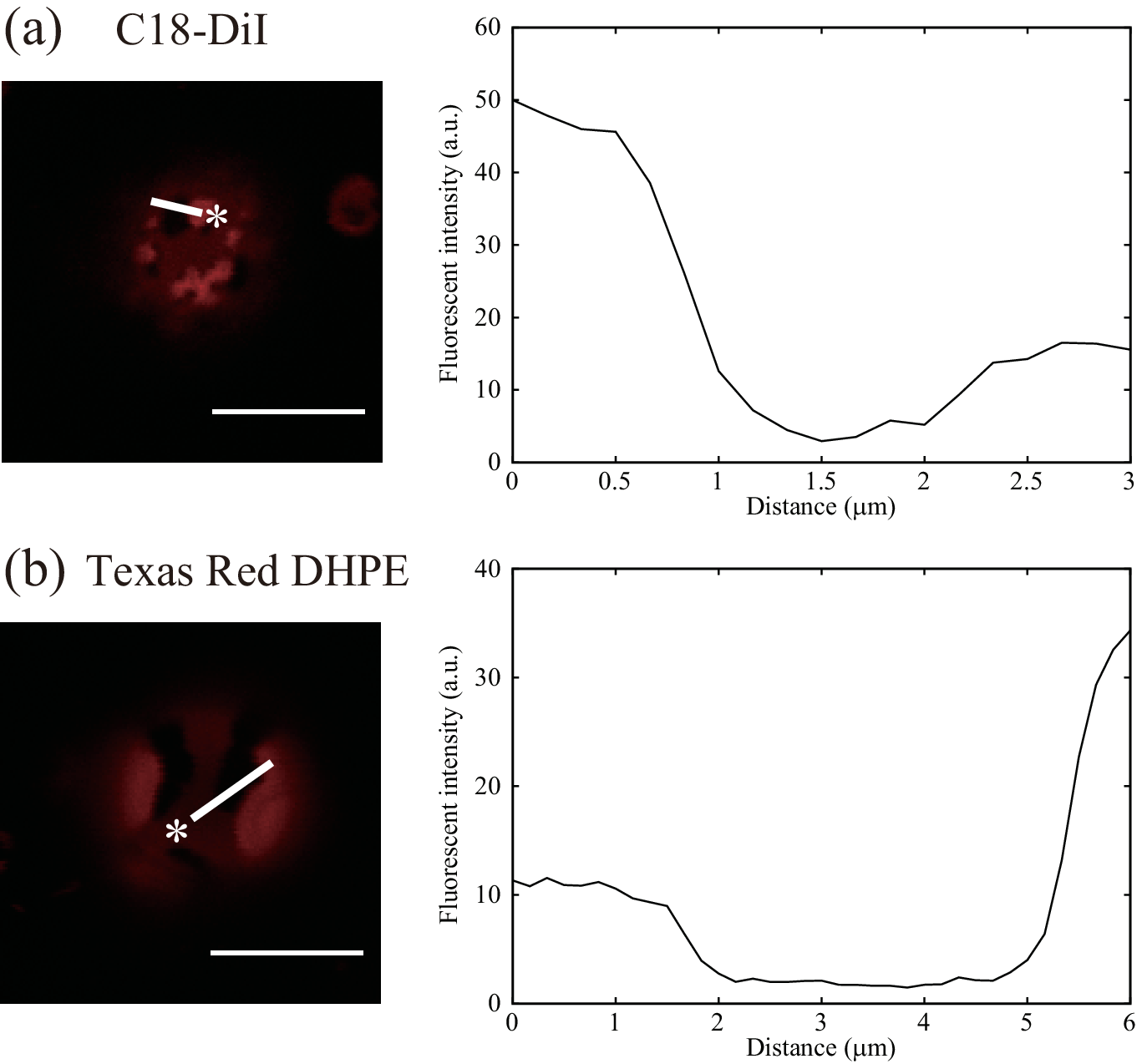}
\end{center}
\caption{
Fluorescence microscopy images obtained after labeling with (a) 1,1'-dioctadecyl-3,3,3',3'-tetramethylindocarbocyanine perchlorate (C18-DiI, PromoKine) and (b) 1,2-dihexadecanoyl-sn-glycero-3-phosphoethanolamine, triethylammonium salt (Texas Red DHPE, Thermo Fisher Scientific). These images were obtained for DOPS/DPPC = 30/70 and $\Delta c$ = 100 mM at pH = 7. The plots on the right show the fluorescence intensity profiles along the white lines from the edges indicated with asterisks in the images. Three phases could be distinguished in (a), namely, regions with high fluorescence intensity (distance from the asterisk: 0 - 1 $\mu$m), low fluorescence intensity (1 - 2 $\mu$m), and intermediate fluorescence intensity (2 - 3 $\mu$m). Similarly, three phases were also observed in (b), namely, regions with intermediate fluorescence intensity (0 - 2 $\mu$m), low fluorescence intensity (2 - 5 $\mu$m), and high fluorescence intensity (5 - 6 $\mu$m). As the three-phase coexistence was also observed using these additional fluorescent probes, the nature of the fluorescent probes did not affect the occurrence of the three-phase coexistence.}
\end{figure}

\begin{figure}[th]
\begin{center}
\includegraphics[scale=0.9]{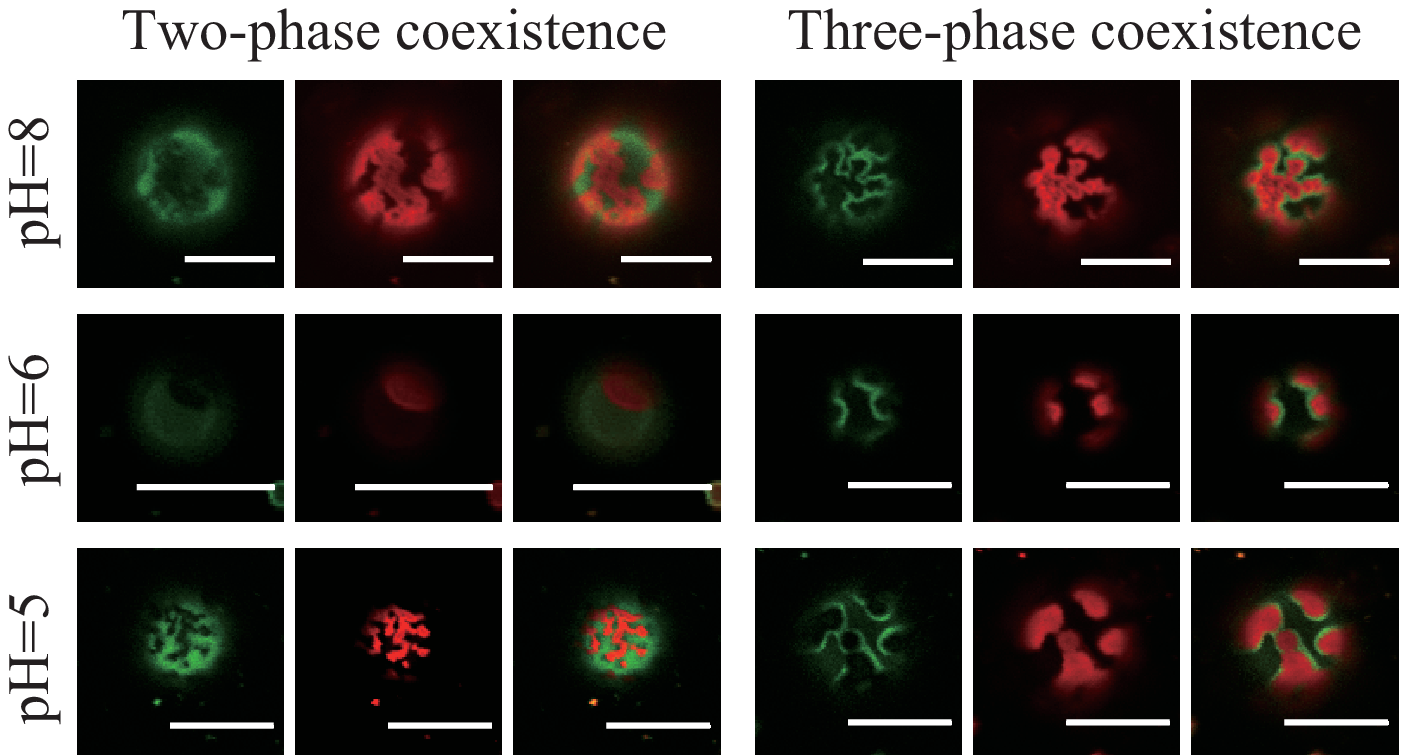}
\end{center}
\caption{
Typical confocal laser scanning microscopy images at $\Delta c$ = 100 mM for DOPS/DPPC = 30/70 at pH = 8, 6, and 5. Scale bars are 10 $\mu$m.}
\end{figure}

\clearpage

\section*{Steady-state fluorescent anisotropy measurement}
In order to monitor the ordering of hydrocarbon tails of DOPS depending on pH changes, we measured the steady-state fluorescent anisotropy of 1,6-diphenyl-1,3,5-hexatriene (DPH) in DOPS-single-component membranes.
DOPS and DPH were dissolved in chloroform to afford concentrations of 20 mM and 0.2 mM, respectively.
260 $\mu$L of DOPS solution and 120 $\mu$L of DPH solution were mixed and the mixed solution was evaporated under a flow of nitrogen gas.
The lipids were further dried in a vacuum desiccator for at least 3 h.
200 $\mu$L of Milli-Q water or pH-adjusted solution was added to the lipid films.
The lipid solution was mixed using a vortex mixer for several seconds and then sonicated at room temperature for 1 h.
800 $\mu$L of MilliQ-water or pH-adjust solution was further added to the lipid solution and then the total amount became 1000 $\mu$L.
The lipid solution was extruded through polycarbonate membranes (pore size is 0.4 $\mu$m, Whatman) in a mini-extruder (Avanti Polar Lipids).
After extrusion for 20 times through a polycarbonate membrane, we took 950 $\mu$L of the sample solution.

100 $\mu$L of the sample solution and 1900 $\mu$L of water of pH-adjust solution were mixed and set in a spectrofluorometer (FP-6500, Jasco, Japan) for the steady-state fluorescent anisotropy measurement.
Samples were excited at $\lambda_{\rm ex}=$ 357 nm and the fluorescent intensity was monitored at $\lambda_{\rm em}=$ 430 nm.
The fluorescent anisotropy $r$ was calculated by the following equation,
\begin{eqnarray}
r = \frac{I_{\rm VV}-GI_{\rm VH}}{I_{\rm VV}+2GI_{\rm VH}} \\
G = \frac{I_{\rm HV}}{I_{\rm HH}}
\end{eqnarray}
where $I$ was the fluorescence intensity.
The subscripts indicated the orientations of polarizers.
First and second characters meant the orientations of excitation and emission polarizers respectively, and H and V meant horizontal and vertical orientations.
Therefore, HH is both polarizers in horizontal orientations, VV is both polarizers in vertical orientations, HV is the excitation polarizer in horizontal orientation and the emission polarizer in vertical orientation, and VH is the excitation polarizer in vertical orientation and the emission polarizer in horizontal orientation.
The fluorescence anisotropy $r$ was measured in increments of 2 $^{\circ}$C from 20 to 35 $^{\circ}$C (20, 22, 24, 26, 28, 30, 32, 34, and 35 $^{\circ}$C), each measurement was performed three times. The results were summarized in Fig.~S9.

\begin{figure}[th]
\begin{center}
\includegraphics[scale=0.7]{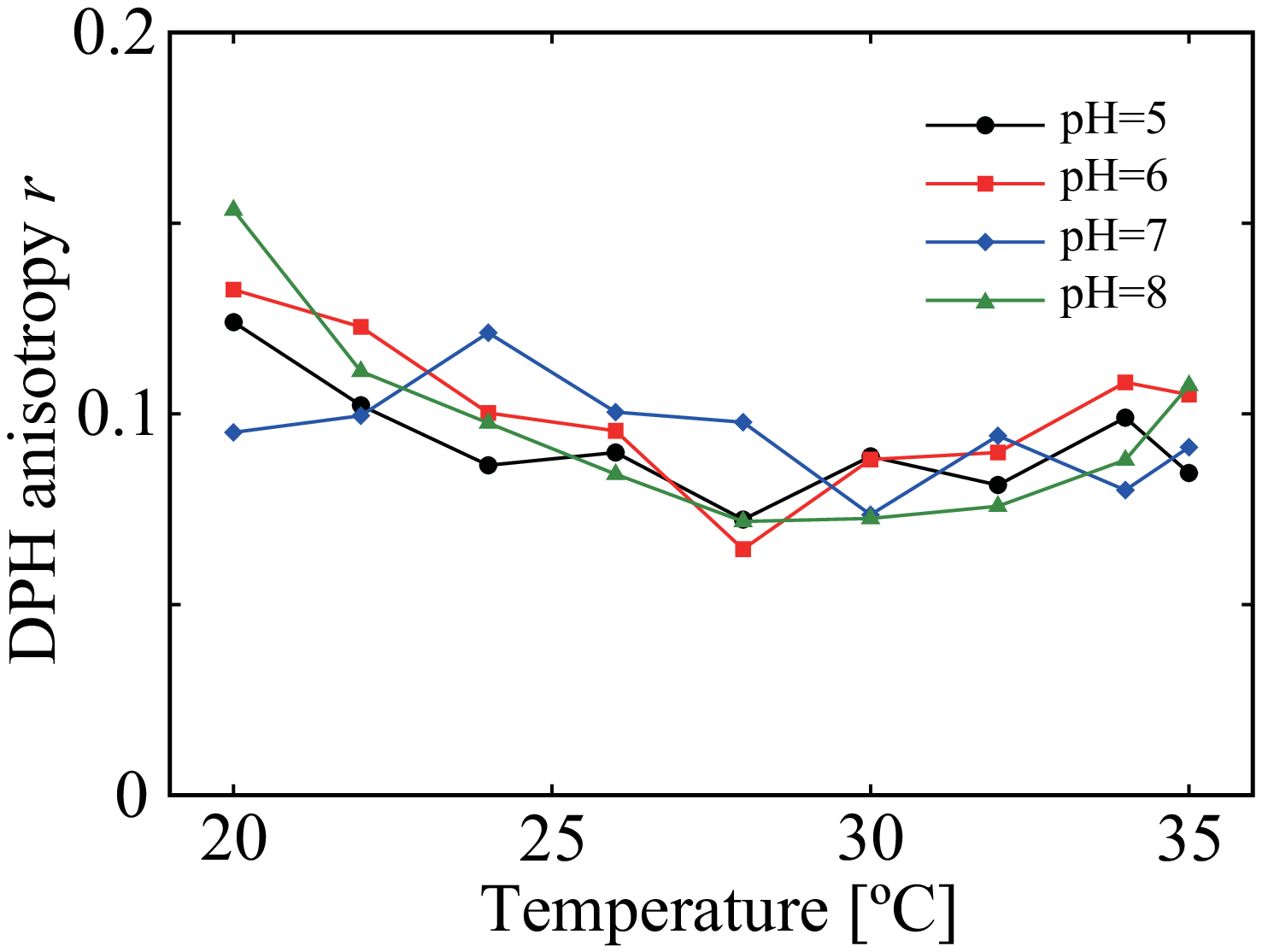}
\end{center}
\caption{
DPH fluorescence anisotropy $r$ (average values of three independent measurements) for pH = 5 (black line with circle symbols), 6 (red line with square symbols), 7 (blue line with diamond symbols), and 8 (green line with triangle symbols). 
Generally, disordered phase has $r \simeq 0.1$ and ordered phase has $r \simeq 0.3$. Even if the pH was changed, there was almost no change in the ordering of the hydrocarbon tails of DOPS, which is a disordered state.}
\end{figure}


\end{document}